\documentclass[aps,prl,showpacs,twocolumn,superscriptaddress]{revtex4}
\usepackage{graphicx}
\usepackage{amsmath}
\usepackage{amssymb}
\bibstyle{apsrev.bib}

\newcommand{\beq}{\begin{equation}}
\newcommand{\eeq}{\end{equation}}

\newcommand{\maya}[1]{{\large{$\clubsuit$}}{\em #1}}
\newcommand{\marco}[1]{{{$\bigstar$}}{\em #1}}
\newcommand{\comment}[1]{}
\newcommand{\change}[2]{{\comment{#1}}{\bf #2}}

\begin{document}

\title{Inter-occurrence Times in the Bak-Tang-Wiesenfeld Sandpile
Model: A Comparison with the Turbulent Statistics of Solar Flares}
\date{\today}
\author{Maya Paczuski}
\email{maya@ic.ac.uk}
\affiliation{Perimeter Institute for Theoretical Physics, Waterloo, Canada, N2L 2Y5}
\author{Stefan Boettcher}
\affiliation{Department of Physics, Emory University, Atlanta, GA 30322, USA}
\author{Marco Baiesi}
\affiliation{Instituut voor Theoretische Fysica, K.U.Leuven, B-3001, Belgium}
\begin{abstract}
A sequence of bursts observed in an intermittent time series may be
caused by a single avalanche, even though these bursts appear
 as distinct events when  noise
and/or instrument resolution impose a detection threshold.  In the
Bak-Tang-Wiesenfeld sandpile, the statistics of
quiet times between bursts switches from Poissonian to scale invariant
on raising the threshold for detecting instantaneous activity, since each
zero-threshold avalanche breaks into a hierarchy of correlated bursts.
Calibrating the model with the time resolution of GOES data,
qualitative agreement with the inter-occurrence time statistics of solar
flares at different intensity thresholds is found.
\end{abstract}


\maketitle
 A simple picture
of intermittency in turbulent flows places rare, hot regions
that dissipate energy inside a cold laminar sea.  As the
hot regions evolve, they
maintain a clustered structure,  and the  dissipation, occurring at
small scales, remains correlated at large ones. For instance, examining the sky at
night, one sees stars, galaxies and clusters of galaxies against a
dark background~\cite{ozernoy,bak_pac_universe}. Many  striking
examples of turbulent intermittency occur in astrophysical, space or laboratory
plasmas, such as flaring events in the solar
corona~\cite{aschwanden:TRACE,lh}, magnetic
substorms~\cite{chang_1999}, bursty bulk flows~\cite{angelopoulous}, auroral emissions~\cite{klimas}, turbulence in the solar
wind~\cite{freeman00,voros}, or bursts observed in 
RFX~\cite{spada,antoni} and 
RFP experiments~\cite{marrelli05:_reduced}.
 Important parameters include the Reynolds number(s), $R=Vl_0/\nu$ (or
$R_m=Vl_0/\eta$, etc.), where $\nu$ is the viscosity, $\eta$ is the
magnetic diffusivity, and $V$ is the velocity difference over the
integral scale $l_0$.
 For increasing $R$, 
characteristic widths of the dissipating regions decrease, while their intensity
increases.
Thus turbulence becomes an
``on/off'' phenomena as the appropriate Reynolds number(s) become
large~\cite{frisch,bak_pac_universe}.

The controversial hypothesis that turbulent intermittency may be a
manifestation of self-organized criticality (SOC) has been discussed
by Bak and
others~\cite{btw,lh,chang_1999,forest_fire,stella,sreenivasan,bak_pac_universe,pac_hughes1,Charbonneau:ModelReview,sanchez02:_wait-SOC,Wheatland:Waiting}. In
this scenario, intermittent energy dissipation is 
 a stick-slip or threshold process.
   Each slip can trigger further slips -- either through short
or long range interactions.  Eventually, a regime materializes
where sparse, sporadic avalanches of intense energy dissipation
interrupt laminar regions of space-time which rest near static
equilibrium with little dissipation despite the continuous, global
input of energy.

Up to now, SOC has mostly been studied with vividly plain models, such
as the original Bak-Tang-Wiesenfeld sandpile (BTW), that ignore many features of hydrodynamic, plasma or other kinds of
 turbulence.
Rather, they generate patterns of rapid energy dissipation in space
and time, which is the hallmark of intermittency. On the other hand,
most studies of turbulence~\cite{frisch} examine
structure functions and multiscaling
phenomena~\cite{she_leveque,dubrulle}, and/or nonlinear instabilities
and coherent structures, etc.~\cite{metais,cowley,chang_1999}, all of
which are associated with intermittency but do not directly
characterize the bursts of energy dissipation in space and time. 
As a result, some comparisons that have been made are 
superficial and should be made more definitive. Further, one argument used so far to distinguish SOC from turbulence is
misleading and erroneous.

While  BTW and other SOC models  exhibit 
a broad distribution of avalanche sizes and durations, which are
comparable to e.g. solar flare data, a marked difference has been
noted regarding the time intervals between bursts.   For instance, 
Boffetta {\it et al.}~\cite{boffetta} found that the
distribution of times between  flares exhibits power law
statistics, while intervals between subsequent avalanches in BTW are (approximately)
Poissonian. Further, they and 
other groups~\cite{freeman00,spada,antoni,carbone,marrelli05:_reduced} 
found that shell or reduced magnetohydrodynamics (MHD)  models gave a better description of the
waiting time statistics, and some~\cite{boffetta,spada,antoni,carbone} used this to
distinguish SOC from turbulence, or to question the applicability of the
SOC paradigm for magnetically confined plasmas in thermonuclear research~\cite{spada}.

We make three key points. First, {\it intermittent
bursts can never be detected, nor distinguished
from the background, at arbitrarily low thresholds.} For
instance, the studies above comparing reduced MHD or shell models with flares
or bursts in man-made plasmas use a threshold for defining bursts.
 Such a threshold is realistic because the emission associated
with e.g. flares decays slowly after a local peak, allowing overlaps with
subsequent peaks.  Although a threshold is unavoidably connected to the
precise definition of events, robust features may be 
observed with rescaled distributions measured at different 
thresholds~\cite{BPS,alvaro1}.   For such time series,
{\it event durations and quiet times between them are
measured on a single clock with equal precision}. 
Thus one can consider the
hypothesis that {\it the sequence of bursts arises from a single avalanche
observed at finite detection threshold.}
The bursts within an avalanche in a SOC system can be expected to be
 correlated in space and time, being part of the same, critical process.

\begin{figure}[!tb]
\includegraphics[angle=0,width=8.0cm]{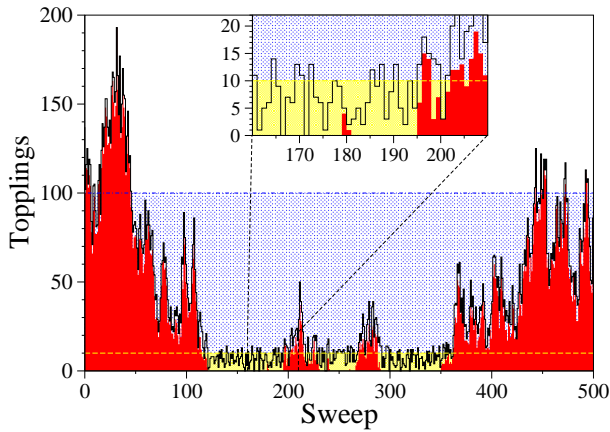}
\caption{(Color online) Time series of the instantaneous number of
topplings, $n(t)$, in the two dimensional BTW  with $L=2048$ and $\Delta T=100$. 
Dark (red online) areas and the black line represent the actual signal and
the signal with some additive noise, respectively. Shaded regions indicate a
detection thresholds  $n_c=10$ (yellow, dashed line) and $n_c=100$ (blue, dot-dashed line); any
signal below $n_c$ is considered undetectable.  Intervals between a
rise of the signal above $n_c$ and its next fall below
constitute measured durations $t_d$ of events, followed by a ``quiet
time'' $t_q$ until the next rise. 
The ``waiting time'' between two
consecutive rises is $t_w=t_d+t_q$.  
\label{fig:alice}}
\end{figure}

To test this idea, we study a BTW sandpile~\cite{btw} in which the
time unit is that of a parallel update step, and consider both slow driving
(A)
and running sandpile~\cite{running-sandpile,sanchez02:_wait-SOC,corral} (B) conditions.
In the time series of the activity, $n(t)$, bursts
are defined as consecutive intervals during which $n(t)>n_c$, where
the detection threshold for events is $n_c\geq 0$ (see Fig.~\ref{fig:alice}).  A data analysis
technique analogous to that recently developed by Baiesi {\it et
al.}~\cite{BPS} to examine inter-occurrence statistics of  flares 
provides a direct comparison between results from numerical simulations of BTW and solar flares.
 It turns out that  BTW
  shares several features with the intermittent statistics of flares,
as shown below. 

The BTW sandpile consists of a $L\times L$ lattice with a discrete
number $z_i$ of sand grains occupying each site~$i$.  We study two versions:
(A) in the  slow driving limit, a grain is added to the pile
at a randomly chosen location
when the previous ($n_c=0$) avalanche ends.   The durations ($t_d$) and quiet times ($t_q$)
then refer to intervals between local peaks within each avalanche and statistics are
obtained over many avalanches. In the second case (B),  one grain of sand is dropped every $\Delta
T$ update steps at a randomly chosen site. In both cases, at each update step, $t$,
all sites that exceed a threshold for stability, $z_i>z_c = 3$, topple in
parallel
by distributing a single grain of sand to each of their four nearest
neighbors or, for boundary sites, over the edge of the lattice.
Taken as the instantaneous dissipation signal, the activity $n(t)$ is
the number of unstable sites toppling at each  parallel update step.
In model~B, if $\Delta T> \langle t_d\rangle_L$,
 the sequence of topplings is also interrupted by instances where the
activity completely stops ($n(t)=0$)~\cite{corral}. The quantity $\langle
t_d\rangle_L$ is the average duration of avalanches on a
lattice of scale $L$ in the stationary state of 
model~A.  For both A and B, consecutive
stopping points separated by intervals where $n(t)>0$ delimit $n_c=0$ 
avalanches. 
The time series of Model B has similar character to the solar flare
data studied in Ref.~\cite{BPS}, with a broad distribution of events
that exceed each threshold $n_c$, albeit with a finite-size cutoff
curtailing the power-law tail observed in Fig.~2a of Ref.~\cite{BPS}.

Consider an observer who measures the global activity sequence with a
finite error, so that the time series she records is $n_{\rm
obs}(t)=n(t) + \eta(t)$. For instance, let $\eta(t)$ be an independent
random number uniformly distributed between 0 and 15. The effect of this noise is shown in
Fig.~\ref{fig:alice} for model~B. 
 One way for the observer to separate the signal from the noise is to
increase her threshold for detecting events, and coarse-grain her unit
of measurement. In observing natural phenomena, such as flares, these detection thresholds are an intrinsic and unavoidable part of the
measurement. Hence, we study the original time
series together with a finite threshold $n_c>0$ (e.~g. $n_c=100$ as in
Fig.~\ref{fig:alice}) to distinguish bursts, considering all instances
with $n(t) \leq n_c$ to have no activity. For sufficiently large
$n_c$, the event statistics that the observer measures (at large
times) are the same as the actual statistics without
noise.

\begin{figure}[!tb]
\includegraphics[angle=0,width=8.0cm]{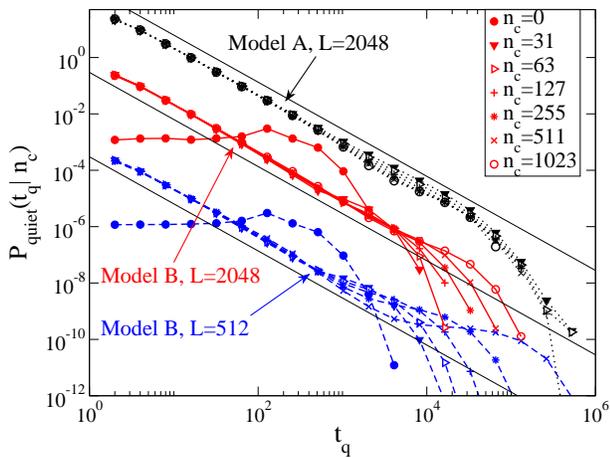}
\caption{(Color online) Distribution of quiet times $t_q$ in  BTW for
  different thresholds $n_c$ for
model~A and  model~B with $\Delta T=100$ (groups of curves are offset vertically). In model B the distribution changes from
approximately exponential at $n_c=0$ to a power law for increasing
$n_c$. The straight reference lines decay
as $t_d^{-5/3}$, suggesting $\gamma_q^{BTW}=1.67(5)$ in
Eq.~(\protect\ref{gammaBTWeq}).  
\label{Pquietplot}}
\end{figure}

As Fig.~\ref{Pquietplot} shows, on increasing the threshold $n_c$ from
zero, the distribution of quiet times $t_q$ for model~B switches from an
(approximately) exponential distribution to a power law,
\beq 
P_{quiet}(t_q) \sim t_q^{-\gamma_q}{\rm \ \ \
with \ \ \ } \gamma_q^{BTW} = 1.67 \pm 0.05.
\label{gammaBTWeq}
\eeq
For model~A one observes an even cleaner and broader 
scaling regime, with the same $\gamma_q^{BTW}$~\footnote{
Error bars quoted here and elsewhere in the text represent our
estimate of statistical and systematic errors associated with finite
size  effects.}.
In model~A, the 
power law tail arises from the correlations 
of bursts within each avalanche. 
Further, the Abelian property of BTW assures that the
power law behavior of quiet times in model~B
 cannot be due to overlapping avalanches.
Thus, the natural introduction of detection thresholds leads to the discovery
of a hierarchical sequence of correlated bursts (or sub-avalanches) within a large avalanche.

\begin{figure}[!tb]
\includegraphics[angle=0,width=8.0cm]{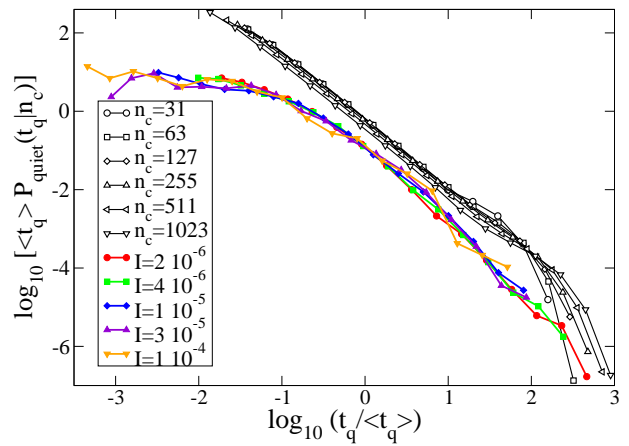}
\caption{(Color online) Rescaled distribution of quiet times in BTW
 with $L=2048$, $\Delta T=100$
  for different thresholds $n_c$ (black lines, open
 symbols) and for the solar flare data corresponding to ``ALL'' at
 different thresholds $I$ in Fig.~3(b) of Ref.~\protect\cite{BPS}
 (colored lines, full symbols). The quiet times $t_q$
 have been divided by the average quiet time $\langle
 t_q\rangle$ at each threshold, and the distributions rescaled
  to preserve normalization. BTW and  flare data are
 similar in the intermediate regime. The former are shifted up by one unit
 on the log scale.
\label{solarBTW}}
\end{figure}

A similar data analysis by Baiesi {\it et al.}~\cite{BPS} for  flares,
detected as intervals during which the emission intensity in the GOES
time series exceeds $I$, used a particularly simple scaling ansatz
for the quiet time distribution,
\beq
P_{quiet}(t_q | I) = \frac{1}{\langle t_q \rangle_I}
f_{flare}\left(\frac{t_q}{\langle t_q \rangle_I}\right).
\label{Pqscaleq}
\eeq
Here $f_{flare}$ is a scaling function and the average quiet time for
a given threshold, $\langle t_q \rangle_I$, provides a
factor that collapses distributions measured at different
$I$.  While BTW
 does not  obey this
particular scaling ansatz, for comparison we apply Eq.~(\ref{Pqscaleq}) to BTW
with different values of $n_c$. This enables us to study the
dependence of two dimensionless quantities $\langle t_q\rangle
P_{quiet}$ vs. $t_q/\langle t_q\rangle$ at different thresholds for
both flares and BTW.  As shown in Fig.~\ref{solarBTW}, we find
that although the scaling functions $f(x)$ are clearly different for
$x \ll 1$, they are similar for intermediate arguments.

\begin{figure}[!bt]
\includegraphics[angle=0,width=8.0cm]{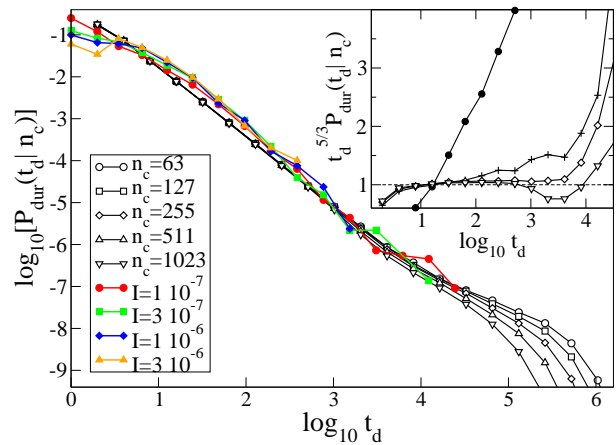}
\caption{(Color online) The distribution of event durations for
 BTW model~B  with $L=2048$, $\Delta T=100$ and for
flare data corresponding to ``minimum'' at different
thresholds $I$ in Fig.~4 of Ref.~\protect\cite{BPS}. One parallel update step in BTW has
been calibrated to the time resolution of the GOES data,
or one minute. Although the critical exponents are different, the overall
distributions are roughly comparable.
The inset shows that,
for increasing $n_c$, the behavior changes from  that in
Refs.~\protect\cite{Lubeck97,stella_2} towards a plateau indicating
an asymptotic exponent $\approx 5/3$ for BTW at large thresholds [$n_c$: 0 ($\bullet$),
15 ($+$), 255 ($\diamond$), and 1023 ($\triangledown$)]. Similar behavior
is found for model~A (not shown).
\label{solarBTWdurplot}}
\end{figure}

Event durations, $t_d$, observed
at different thresholds in BTW and for flares occurring
during solar minimum, are shown in Fig.~\ref{solarBTWdurplot}~\footnote{For
flares, the distribution of burst durations is
independent of threshold, but depends on the phase of the solar
cycle~\cite{BPS}.}.  In order to compare with BTW, we set
the time resolution of the  GOES  data equal to the time
resolution of BTW. One minute is set equal to one
sweep. Using this straightforward
calibration, the statistics of event durations are similar as well.
 In both cases, power law behavior is
observed, $P_{dur}(t_{d}) \sim t_{d}^{-\gamma_{dur}}$.  
For flares at solar minimum $\gamma_{dur}=2.0 \pm 0.1$~\cite{BPS}, while
this exponent the $n_c \gg 1$ bursts in BTW is
smaller.

The inset of Fig.~\ref{solarBTWdurplot} shows how the apparent power law behavior for
burst durations in  BTW changes with increasing threshold $n_c$.
At sufficiently large $n_c$ a plateau appears in the function
$t_{d}^{5/3}P_{dur}(t_{d})$. This suggests a critical exponent at high
thresholds, $\gamma_{dur}^{BTW}=1.67 \pm 0.05$, similar to
$\gamma_q^{BTW}$.
 The figure also demonstrates that the scaling behavior of
burst durations, for the system sizes that were studied,  
differs at large $n_c$ from that at $n_c=0$.
In the latter case, L\"ubeck and Usadel~\cite{Lubeck97}
determined a critical exponent $\gamma_{dur}^{BTW}(n_c=0) \simeq 1.48$,
while Stella and De Menech~\cite{stella_2} found multiscaling.

If the sequence of quiet times $(t_q)_i$ are uncorrelated, the cumulative variable
$y_l(j)= \sum_{i=j}^{i=j+l}(t_{q})_i$ exhibits diffusive
behavior, with an average variance scaling as $\sigma= (\langle
y^2\rangle - \langle y\rangle^2)^{1/2} \sim l^{H}$, with $H=1/2$~\cite{demenech:waves-aval}.
Our measurements of this quantity confirm that quiet time
intervals for BTW at different thresholds are uncorrelated,
in all cases giving $H=1/2$.  However, at solar minimum, 
we get $H \simeq 0.62$ (see also~\cite{leddon}).  Therefore,
BTW does not reproduce  correlations {\it between} quiet times for flares.

In fact, a variety of SOC models exhibit power laws in times between
events.
Previous analyses~\cite{norman01, hedges05:_OFC_waiting, baiesi05} 
using a detection threshold considered a completely
different limit than that discussed here, namely that of an infinite time
scale separation between the driving rate and the durations of  events.
Consequently, a threshold was imposed on a variable related to the total
size or
duration of each $n_c=0$
 avalanche, rather than its instantaneous dissipation.
However, as explained earlier our limit of overlapping time scales 
for durations and quiet times may be the correct one to describe
intermittency in turbulence. 
Indeed, cellular automata models of laboratory plasmas are 
running sandpiles; see e.g. Newman {\it et al}~\cite{newman96}.
Within this scheme, some works~\cite{sanchez1,sanchez02:_wait-SOC}, 
also using a threshold, have found power law quiet times
for the Hwa-Kardar~\cite{running-sandpile} 
running sandpile driven at a sufficiently high rate.
This was claimed to be due to interactions between
 overlapping  avalanches. In contrast,
here we show that the BTW model, considering one avalanche at a time, generates
a power law distribution of quiet times when a finite detection threshold
is used. Its Abelian property assures that this correlation remains
the same when the model is driven at a finite rate, in agreement with the
results shown here. It is unlikely, though, that the Abelian property of
 BTW is
essential to getting a scale free distribution of quiet times, although this
remains to be clarified by studying other models in a similar way.

In conclusion, we have demonstrated that SOC remains a viable
 alternative for the explanation of intermittent dissipation in
 turbulence by comparing results from numerical simulations of a  BTW
 sandpile with solar flare statistics. By including an inevitable
 detection threshold in the analysis of  BTW, as well as allowing an
 overlap of time scales between burst durations and quiet times,
 qualitative correspondence is obtained for the power law statistics
 of inter-occurrence times for solar flares.   
 Studies of more
 physically realistic SOC including detection thresholds for short
 time dissipation in the whole system may improve quantitative
 agreement.

M.~P. thanks Peter Grassberger for comments on the manuscript. 
M.~B.~acknowledges support from an FWO
post-doctoral position (Flanders). S.~B.~thanks the Perimeter Institute for
its hospitality.

\bibliographystyle{apsrev}
\bibliography{biblio_btw_flare}

\end{document}